\documentclass[11pt]{article}
\usepackage{amssymb}
\usepackage{amsmath}
\usepackage{graphicx}
\usepackage{color}

\def\beq{\begin{eqnarray}}    
\def\eeq{\end{eqnarray}}      

\newcommand{\CC}{\Lambda}
\newcommand{\rLo}{\rho_{\CC}^0}
\newcommand{\rL}{\rho_{\CC}}
\newcommand{\rmr}{\rho_m}

\newcommand{\pmr}{p_m}
\newcommand{\w}{\omega}

\newcommand{\ba}{\begin{eqnarray}}
\newcommand{\ea}{\end{eqnarray}}
\newcommand{\brr}{\begin{array}}
\newcommand{\err}{\end{array}}
\newcommand{\bc}{\begin{center}}
\newcommand{\ec}{\end{center}}

\newcommand{\be}{\begin{equation}}
\newcommand{\ee}{\end{equation}}

\newcommand{\mysection}[1]{\section{#1}
\renewcommand{\theequation}{\thesection.\arabic{equation}}
\setcounter{equation}{0}}

\begin{document}

\hyphenation{cos-mo-lo-gists un-na-tu-ral-ly ne-cessa-ry dri-ving
par-ti-cu-lar-ly a-na-ly-sis mo-del mo-dels ex-ten-ding e-xam-ples
ho-we-ver}

\begin{center}

{\LARGE \textbf{Nonsingular Decaying Vacuum Cosmology and Entropy
Production}} \vskip 2mm

 \vskip 8mm

 \vskip 8mm

\textbf{J. A. S. Lima}

\vskip0.25cm Departamento de Astronomia, Universidade de S\~ao
Paulo, Ruado Mat\~ao 1226, 05508-900, S\~ao Paulo, Brazil

\vspace{0.25cm} E-mail: jas.lima@iag.usp.br \vskip2mm

\textbf{Spyros Basilakos}

\vskip0.25cm

Academy of Athens, Research Center for Astronomy \& Applied
  Mathematics, \\ Soranou Efessiou 4, 11-527, Athens, Greece

\vspace{0.25cm} E-mail: svasil@academyofathens.gr \vskip2mm

\textbf{Joan Sol\`a}

\vskip0.25cm High Energy Physics Group, Dept. d'Estructura i
Constituents de la Mat\`eria, and Institut de Ci\`encies del Cosmos
(ICC), \\ Univ. de Barcelona, Av. Diagonal 647,
 E-08028 Barcelona, Catalonia, Spain

\vspace{0.25cm} E-mail: sola@ecm.ub.edu \vskip2mm

\end{center}
\vskip 15mm


\begin{abstract}
The thermodynamic behavior of a decaying vacuum cosmology describing
the entire cosmological history evolving between two extreme (early
and late time) de Sitter eras is investigated. The  thermal
evolution from the early de Sitter to the radiation phase is
discussed in detail. The temperature evolution law and the
increasing entropy function are  analytically determined. The
entropy of the effectively massless particles is initially zero but
evolves continuously to the present day  maximum value  within the
current Hubble radius, $S_0 \sim 10^{88}$ in natural units. By using
the Gibbons-Hawking temperature relation for the de Sitter
spacetime, it is found that the ratio between the primeval and the
late time vacuum energy densities is $\rho_{vI}/\rho_{v0}  \sim
10^{123}$, as required by some naive estimates from quantum field
theory.

\vskip0.2cm

\end{abstract}

\mysection{Introduction} The possibility of a nonsingular early de
Sitter phase driven by a decaying vacuum was  phenomenologically
proposed to solve several problems of the Big-Bang cosmology (Lima
\& Maia 1994; Lima \& Trodden 1996).  It traces back to some early
attempts to solve (or at least to alleviate) several cosmological
puzzles, in particular the  ``graceful exit" problem plaguing  most
variants of inflation (for recent reviews Guth, Kaiser  \& Nomura
2013; Linde 2014). Difficult and arduous as all these theoretical
conundrums are, the most serious one is the cosmological constant
problem (Weinberg 1989; Sahni \& Starobinsly 2000; Padmanabhan 2003;
Peebles \& Ratra 2003; Copeland, Sami, Tsujikawa 2006), which
remains as the main challenge also for inflationary theories of the
cosmic evolution.

Later on, new developments have appeared which could mitigate some
of these problems. For instance, by combining the phenomenological
approach with the renormalization group (RG) theoretical techniques
of quantum field theory (QFT) in curved spacetimes, a large class of
dynamical $\Lambda$(H)-models  described by an even power series of
the Hubble rate was proposed (Shapiro \& Sol\`a 2002, Sol\`a 2008,
Shapiro \& Sol\`a 2009).

More recently, we have proposed a large class of nonsingular
cosmologies providing a complete cosmological history  evolving
between two extreme (early and late-time) de Sitter phases supported
by a dynamical decaying vacuum energy density (Lima, Basilakos \&
Sol\`a 2013;  Basilakos, Lima \&  Sol\`a 2013; Perico et al. 2013;
Sol\`a \& G\'omez-Valent 2015). Unlike many inflationary variants
which are usually endowed with a preadiabatic phase, this
nonsingular decaying vacuum is responsible by an increasing entropy
evolution since the start of the primeval de Sitter phase.

The consistency and constraints imposed by the generalized second
law (GSL) on this class of nonsingular cosmologies, including the
matter-energy content and the horizon entropy of both de Sitter
phases, were investigated (Mimoso \& Pav\'on, 2013). The model
approaches thermodynamic equilibrium in the last de Sitter era in
the sense that the entropy of the apparent horizon plus that of
matter and radiation inside it increases and is concave. It was also
found that  the consistency with the GSL is maintained even  when
quantum corrections to the Bekenstein-Hawking entropy are
considered.

Here, instead to analyze the general thermodynamic conditions that
must be obeyed in principle by such spacetimes, we focus our
attention on  the CMB entropic content  generated by this kind of
$\Lambda(H)$ nonsingular cosmology. We know that the amount of
radiation entropy contained within our horizon, $S_{r0}\sim 10^{88}$
(in natural units),  is lesser than the current estimates of
supermassive black holes, $S_{BH} \sim 10^{103}$,  and much lower
than the holographic bound, $S_{H} \sim 10^{123}$ (Frampton \&
Kephart 2008, Frampton et al. 2009). However, it is still enormous
and well quantified, and this fact creates severe theoretical
problems since we cannot trace back the origin of that entropy
within the conventional description of the cosmic evolution. Perhaps
more interesting, in our model the very early and late time Universe
have the same  nonsingular de Sitter structure which seems to be in
agreement both with the old suggestion  of an early de Sitter phase
supported by quantum corrections for the general relativity
(Starobinsky 1980, Vilenkin 1982),  as well as with the present
observed `quasi' de Sitter Universe.

In the picture explored here the universe starts from an unstable
vacuum de Sitter phase with no matter or radiation. The decay of
this primeval vacuum state generates all the energy (and entropy) of
the matter-energy content. At the end of the process, an extremely
small  remnant of vacuum energy survives.  As we shall see, the
continuous  generation of entropy is very large and its final value
coincides with the present radiation entropy existing within the
Hubble radius. The early decaying vacuum process is not plagued with
the inflationary  ``exit problem" and generates the correct number,
$S_0 \simeq 10^{88}$ (Kolb \& Turner 1990). In addition,  the ratio
between the primeval and the present day vacuum energy densities is
$\rho_{vI}/\rho_{v0}  \simeq 10^{123}$.

The paper is organized as follows. In section 2 we provide a brief
discussion of the dynamical $\Lambda(H)$ model and in section 3 we
set up the basic equations driving its complete evolution. In
section 4, we discuss the  transition from an early Sitter to the
radiation phase. How inflation ends ``gracefully''  and some
estimates of the corresponding reheating temperature are presented
in section 5, while the entropy production generated by the vacuum
decaying component is addressed in section 6.  Finally, in the last
section the main conclusions are delivered.

\mysection{Minimal model for a complete cosmic history} The simplest
$\Lambda(H)$-scenario accounting for the general chronology of the
universe from inflation until our days is based on the following
expression for the vacuum energy density (Lima, Basilakos \& Sol\`a
2013, from now on LSB paper):

\begin{equation}\label{powerH}
\Lambda(H)= c_0+3\nu H^2 +3\alpha\frac{H^4}{H_I^2}\,,
\end{equation}
where $H$ is the Hubble rate and  by definition
$\rL(H)=\CC(H)/8\pi\,G$ is the corresponding vacuum energy density
($G$ being Newton's constant). The even powers of $H$ were selected
due the general covariance of the effective action as required by
the QFT treatment in curved spacetime (Shapiro \& Sol\`a 2002 and
2009; Sol\`a 2008 and 2013).

The constant $c_0$ in the above expression yields the dominant term
at low energies, when $H$ approaches the measured value $H_0$ (from
now on the index ``0" denotes the present day values).  The
dimensionless parameter $\nu$ can be determined from observations
(Espa\~na-Bonet et al. 2004; Basilakos, Plionis \& Sol\`a 2009;
Grande  et al. 2011; Basilakos, Polarski \& Sol\`a 2012\footnote{See
(G\'omez-Valent \& Sol\`a, 2014; G\'omez-Valent, Sol\`a \& Basilakos
2015) for the latest comprehensive analysis.})  from a joint
likelihood analysis involving SNe Ia, Baryonic Acoustic Oscillations
(BAO) and Cosmic Background Radiation (CMB) data, with the result
$|\nu| \equiv {\cal O}(10^{-3})$. It is of course small, since a
dynamical model of the vacuum energy cannot depart too much from the
phenomenologically  successful $\CC$CDM. It is also encouraging that
within a generic Grand Unified Theory (GUT) one can estimate its
value theoretically in the range $|\nu| \sim 10^{-6}-10^{-3}$
(Sol{\`a}, 2008). Finally, the dimensionless $\alpha$-parameter can
be fixed to unity since it can be absorbed by the undetermined scale
$H_I$.

As it appears, such dynamical vacuum model is able to link the fast
dynamics of the primordial vacuum state with the much more relaxed
evolution of the presently observed Universe.  At early times, the
primeval de Sitter phase (supported by a huge pure vacuum energy
density $\rho_{I} = 3H_I^{2}/8\pi G$) is unstable and drives the
model continuously to a late time de Sitter state characterized by
the remnant vacuum energy density today.

\section{$\Lambda(H$) model: Basic Equations}

It is widely known that the bare cosmological constant appearing on
the geometric side of the Einstein field equations (EFE)  can be
absorbed  on the the matter-energy side  in order to describe the
effective vacuum contribution with energy density
$\rho_{\Lambda}=\Lambda/(8\pi G)$  and pressure
$p_{\Lambda}=-\rho_{\Lambda}$. Hence, by assuming a spatially flat
FLRW (Friedmann-Lema\^\i tre-Robertson-Walker)  metric,  the EFE can
be written as
\be 8\pi G \rmr+\Lambda (H) = 3H^2\,, \label{friedr} \ee \be
\label{friedr2} 8\pi G \pmr-\Lambda (H) = - 2{\dot H} - 3H^2\,, \ee
where an overdot means cosmic time differentiation and the index $m$
refers to the dominant fluid component (nonrelativistic matter or
radiation). The above expressions are valid irrespective of the
constant or variable nature of $\CC$, provided it only evolves with
the cosmic time, but not with space (as this would violate the
cosmological principle). Thus, in our case, the only difference with
respect to the more conventional field equations is the fact that
$\CC=\CC(H)$. In this framework, the local energy-conservation law
[$u_{\mu} (T_{m}^{\mu \nu} + T_{\Lambda}^{\mu \nu})_{;\nu}  =0$]
reads:
\begin{equation}
\dot{\rho}_{m}+3(1+\w)H\rho_{m}=-\dot{\rho_{\Lambda}}\,, \label{ECL}
\end{equation}
where the equation of state (EoS), $p_m = \w \rho_m$, has been
adopted for the fluid component.
We recall that the above result can also be derived directly from
the EFE, as the expression (\ref{ECL}) is actually a first integral
of the original equations.

Finally, by combining the matter EoS with (\ref{friedr}),
(\ref{friedr2}) and inserting (\ref{powerH}) into the result one
finds:

\begin{equation}
\label{HE}
\dot{H}+\frac{3}{2}(1+\w)H^2\left[1-\nu-\frac{c_0}{3H^2}-\alpha\frac{H^2}{H_I^2}\right]=0\,.
\end{equation}
The most interesting aspect of such cosmology is that it encompasses
in a single unified approach both the inflationary and the current
dark energy phases. Concerning the latter, let us emphasize that
since $|\nu|\ll 1$  at low redshifts the model becomes almost
indistinguishable from the cosmic concordance model. The influence
of the $\nu$-parameter in the current Universe is qualitatively new:
it endows the vacuum energy of a mild dynamical evolution (which
could be observed nowadays and appear as dynamical dark energy). The
low-energy behavior is thus very close to the concordance model, but
it is by no means identical. It could actually lead to distinctive
new features in different domains such as the possible variation of
the fundamental constants (Fritzsch \& Sol\`a 2012, 2014 and 2015;
Sol\`a 2014).

Most conspicuously, at high energies the model (\ref{powerH}) leads
to inflation. The reason is because Eq.\,(\ref{HE}) has the
particular solution $H_p=\sqrt{\frac {1-\nu}{\alpha}} H_I$, which is
only valid in the high energy scope where the term $c_0/H^2\ll 1$ of
(\ref{HE}) can be neglected. That solution obviously corresponds to
an inflationary (de Sitter) phase marking a (nonsingular) starting
point of the cosmic evolution (see LSB for more details).

\mysection{Deflation from the primeval de Sitter stage}

In this and the following sections,  without loss of generality, we
fix $\nu=0$ (which is not essential for the study of the early
universe) and $\alpha = 1$ (since we can rescale the value of
$H_I$). In this case, it is easy to check that the solution of
(\ref{HE}) at early times (i.e., when $c_0$ can be neglected), can
be written as

\begin{equation}\label{HS1}
 H(a)=\frac{H_{I}}{\sqrt{D\,a^{3(1+\w)}+1}}\,.
\end{equation}
Upon integration we find the time evolution:
\begin{equation}\label{eq:afactor} H_It=\int_{a_i}^a\frac{d\tilde{a}}{\tilde{a}}\left[1+D\,\tilde{a}^{3 (1+\omega)}\right]^{1/2}\,.
\end{equation}
Here $t$ is the time elapsed since the scale factor increased from
some arbitrary initial point $a_i$ up to the final one $a(t)$. The
constant $D>0$  will be specified in the next section.

Using Eq.(\ref{HS1}) and integrating the EFE
(\ref{friedr})-(\ref{friedr2}) we  obtain the corresponding energy
densities:
\begin{eqnarray}
\rL(a)=\frac{\CC(a)}{8\pi G}=
\frac{\rho_I}{\left[Da^{3(1+\w)}+1\right]^2}\,,\label{eq:densities1}\\
\rmr(a)=\rho_I\,
\frac{Da^{3(1+\w)}}{\left[Da^{3(1+\w)}+1\right]^2}\label{eq:densities2}\,,
\end{eqnarray}
where $\rho_{I}$ is the energy density associated to the primeval de
Sitter stage:

\begin{equation}\label{rhoI}
\rho_{I}=\frac{3H_{I}^{2}}{8\pi G}\,.
\end{equation}
From Eq.\,(\ref{HS1}) we find that  $H \sim H_I$ while the condition
$Da^{3(1+\w)}\ll 1$ is fulfilled, and from (\ref{eq:afactor}) we
learn that the evolution of the scale factor during this initial
period is exponential: $a(t)\sim a_i e^{\,H_I\,t}$. It means that
the Universe is initially driven by a pure de Sitter vacuum state,
and therefore is inflating.

The above mentioned  de Sitter inflationary phase, however, is only
ephemeral. For as soon as the  condition $Da^{3(1+\w)} \ll 1$ is
relaxed,  Eq.\,(\ref{eq:densities1}) implies that the vacuum energy
density  $\rL(a)$ decays. Not only so, it decays quite fast  because
$\rL\sim a^{-6(1+\w)}$ while $\rmr\sim a^{-3(1+\w)}$. By assuming
that the vacuum decays mainly into relativistic particles ($\w =
1/3$) one finds that $\rL \sim a^{-8}$ and $\rho_{r} \sim a^{-4}$.

In effect,  it is easy to check from Eq.\,(\ref{eq:afactor}) that in
the post-inflationary regime, i.e. for $a\gg a_i$  (with
$Da_i^{3(1+\w)}\gg 1$), we are led to
 $a\sim t^{1/2}$ for $\omega = 1/3$. This confirms that the Universe evolves continuously from inflation towards the standard (FLRW) radiation dominated stage.

The upshot is that the Universe becomes rapidly dominated by
radiation and free from the vacuum energy. The cosmic expansion
tends automatically to approach the conventional FLRW regime. This
means that the  primordial Big Bang Nucleosynthesis (BBN) epoch can
proceed fully standard, with no interference at all from an unduly
large amount of vacuum energy,  thereby preserving all the virtues
of the conventional BBN picture.

In Fig. 1, we display  the early  period of the nonsingular model
described above, putting emphasis on the transition from the the
inflationary epoch to the standard FLRW radiation regime.  More
details on this transition (in particular, the so-called ``graceful
exit'', as well as the temperature evolution) will be discussed in
the next section.

\begin{figure}[t]
  \begin{center}
      \resizebox{0.7\textwidth}{!}{\includegraphics{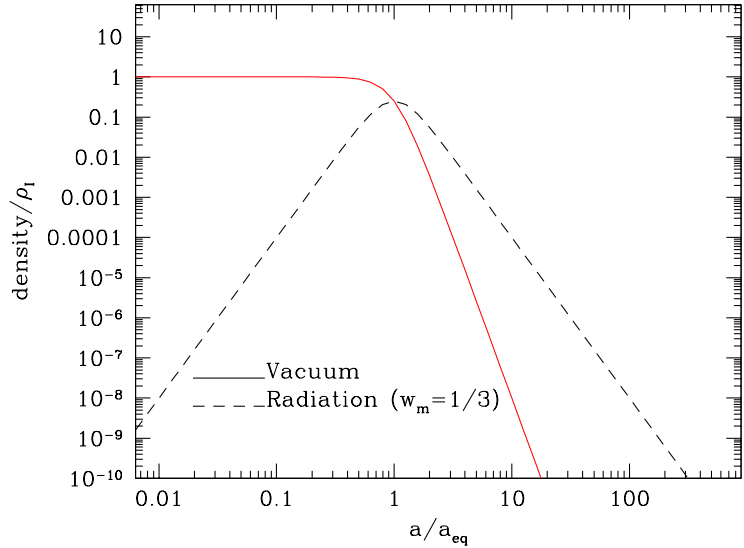}}
      \end{center}
    \caption{Evolution of the
energy densities as a function of the scale factor during the
primeval vacuum epoch (inflationary period) and their transition
into the FLRW radiation era. The densities are normalized with
respect to the primeval vacuum energy density $\rho_I$, and the
scale factor with respect to $a_{eq}$, the value for which
$\rho_{\Lambda} = \rho_r$ (see the text). The curves shown are:
ultrarelativistic particles (black dashed line) and vacuum (red
solid line). See also LSB for more details.}
  \label{Fig1}
\end{figure}

\mysection{The ``graceful exit''}

Let us now demonstrate that in the present nonsingular framework the
inflationary phase (${\ddot a > 0}$) finishes naturally when  the
radiation dominated era starts.  From Eqs.  (\ref{eq:densities1})
and (\ref{eq:densities2}) the exact ratio of the radiation energy
density ($\omega=1/3$) to the vacuum energy density is given by:
\begin{equation}
 \frac{\rho_{r}(a)}{\rho_{\Lambda}(a)} = D a^{4}\,. \label{eq6}
\end{equation}
We denote by $a_{eq}$ the scale factor at the point of
vacuum-radiation ``equality'',  i.e. the precise value marking  the
beginning of the radiation phase at the end of the vacuum-dominated
period\,\footnote{That first equality point $a_{eq}$ in the cosmic
history of our dynamical vacuum universe should, of course, not be
confused with the  usual equality time between the energy densities
of radiation and cold matter, which occurs at redshift $z_{cdm}\sim
3300$
, and hence much later (a factor of $10^{25}$!) in the cosmic
evolution.}. It is obtained from the condition $\rho_{r}(a_{eq})
=\rho_{\Lambda}(a_{eq})$, which renders $D\,a_{eq}^4 = 1$. The last
relation enables us to rewrite the expression (\ref{HS1}) for the
Hubble parameter in the following way:
\begin{equation}
H = \frac{{H}_I}{\sqrt{1 + (\frac{a}{a_{eq}})^{4}}}\,. \label{eq7a}
\end{equation}
It follows that the value of the Hubble function at the
vacuum-radiation equality is simply related to the corresponding
value at the beginning of the de Sitter phase:
\begin{equation}
H_{eq} \equiv H(a_{eq})= \frac{{H}_I}{\sqrt{2}}. \label{eq7b}
\end{equation}
It is also useful to understand dynamically the physical meaning of
the pair ($a_{eq}, H_{eq}$). In principle, we expect that the outset
of the radiation-dominated era should be defined from the condition
that the Universe performs the transition from acceleration into
deceleration, or more precisely when the deceleration parameter
moves from $q=-1$ to $q=1$. From the equation of motion (\ref{HE})
one may check that the decelerating parameter, $q=-{\ddot
a}/aH^{2}=-1-\dot{H}/H^2$, reads:
\begin{equation}
 q(H) = 2 \left[ 1 - \left(\frac{H}{{H}_I}\right)^2 \right] - 1 \,, \label{eq3}
\end{equation}
so that the condition $H=H_{eq}$   implies  that $q=0$ -- see
Eq.\,(\ref{eq7b}).  However,   the start of the radiation phase in
this model is not characterized by the canonical radiation value
$q=1$, as one might have naively expected.

In point of fact, there is  some delay in acquiring this value, that
is, there is an interpolation period during which $q$ moves
continuously from $q = 0$ to the standard result $q = 1$. This
happens because at the equality scale factor the vacuum is still
powerful and continues draining photons in an appreciable rate to
the radiation component (see Fig.\,1).  However, we can estimate
when the standard radiation phase ($q=1$, $H=H_{rad}$, $a =
a_{rad}$) begins  in our model. The decelerating parameter once the
adiabatic radiation regime applies reads:
\begin{equation}
 q(H_{rad}) = 2 \left[1 - \left(\frac{H_{rad}}{{H}_I}\right)^2 \right] - 1\,.  \label{eq15}
\end{equation}

For all practical purposes, we may assume that the radiation phase
has been attained when, say,  $q(H_{rad}) \cong 0.999$. As one may
check from (\ref{eq15}), this happens for $H_{rad}=H_I/(2\times
10^{3})^{1/2}$.  On inserting this value into Eq. (\ref{eq7a}) we
obtain the ratio $a_{rad}/a_{eq} \simeq (2\times 10^{3})^{1/4}$. In
general, by assuming that $q \simeq 1-10^{-x}$ one may show that
$a_{rad}/a_{eq} \simeq (2\times 10^{x})^{1/4}\gg 1$ for  $x\gg1$.
Thus, the scale factor value at the time when the Universe enters in
full the radiation epoch can be one order of magnitude, or more,
larger than the corresponding value at the vacuum-radiation
equality\footnote{Indeed, the precise value of the ratio
$a_{rad}/a_{eq}$  is not important to define the total entropy
generated and other physical quantities in the begin of the
isentropic radiation phase. As we shall see, the important fact is
that a power $n \geq 3$ of such a ratio is much greater than
unity.}. Subsequently, the Universe enters the cold dark matter
[Einstein-de Sitter, $a(t)\propto t^{2/3}$] dominated phase; and,
finally, makes its way to the present vacuum-dust epoch, in which
$\Lambda \simeq \Lambda_{0}=const.$ This will play a role in the
subsequent considerations.

\mysection{Radiation Temperature Law} As shown long ago, when the
vacuum decays ``adiabatically'', that is, in such a way that the
specific entropy of the massless particles remains constant (despite
the total entropy may be increasing), some  equilibrium relations
are preserved (Lima 1996; 1997). In particular, this happens with
the energy density versus temperature relation ($\rho_r \propto
T_r^{4}$) and particle number ($n_r \propto T^{3}$). However, the
temperature does not satisfy the standard scaling law, $Ta =$
constant. In the present scenario, the evolution law for  $T$ can be
determined as follows.

In terms of the scale factor at equality ($a_{eq}$) and using
(\ref{eq:densities2}), the radiation density can be rewritten as:
\begin{equation}\label{eq:rhorrhoII}
\rho_{r} ={\rho}_I \frac{\left(a/a_{eq}\right)^4}{\left[1 +
(a/a_{eq})^{4}\right]^{2}}\,.
\end{equation}

Now, under such  ``adiabatic'' condition the above expression can be
identified with the formula relating $\rho_r$ with the radiation
temperature $T_r$ (see also Lima \& Baranov 2014).  By adding all
the degrees of freedom (d.o.f.) of the created massless modes
through the  $g_{*}$-factor, we may write (Kolb \& Turner 1990):
\begin{equation}\label{eq:rhorKolbTurner}
\rho_{r} =\frac{\pi^2}{30} g_{\ast} T_{r}^{4} \,,
\end{equation}
and combining with (\ref{eq:rhorrhoII}) we obtain for the radiation
temperature law:
\begin{equation}\label{eq:Tr}
T_r = \sqrt{2}T_{rm}\frac{(a/a_{eq})}{\left[1 +
(a/a_{eq})^{4}\right]^{1/2}}\,,
\end{equation}
where $T_{rm}=T_r (a_{eq})$. The maximum value of the radiation
temperature is given by:
\begin{equation}\label{Temp}
T_{rm} = \left(\frac{15\,\rho_I}{2\pi^{2}g_*}\right)^{1/4} =
\left(\frac{45\,H_I^{2}}{16\pi^{3}Gg_*}\right)^{1/4}\,,
\end{equation}
which depends solely on the arbitrary initial scale of the primeval
de Sitter phase ($\rho_I$, or equivalently $H_I$).

The evolution of the temperature of the universe in our model is
shown in Fig.\,2 (top plot). For small values of the scale factor
($a \ll a_{eq}$), the radiation temperature (\ref{eq:Tr}) raises
linearly with the scale factor, $T_r\propto a$, attaining  its
maximum value  at the vacuum-radiation equality ($a=a_{eq}$).
Therefore, instead of having a highly non-adiabatic ``reheating''
event, as typically occurring in the inflaton-like formulations  --
see the conventional picture in e.g. (Kolb \& Turner, 1990) and
reference therein -- in our case we have a relatively long
non-equilibrium heating period in which the vacuum instability
drives progressively the model to the radiation phase. During this
period there is a continuous increase of the temperature and the
entropy (see the next section). As mentioned, the temperature
reaches a maximum given by (\ref{Temp}), and for $a>a_{rad}$ it
decreases in the standard way according to the adiabatic scaling
regime, $T_r \sim a^{-1}$ (cf. Fig.\,2, top).  The linear raising of
the temperature and its subsequent evolution until attaining the
perfect fluid adiabatic  phase is, in our context, directly
responsible for the large radiation entropy observed in the present
Universe.

The following question is in order: {\it Can we calculate the
characteristic quantities ($\rho_I, H_I$) from first principles or
by some reasonable physical condition?} This is indeed a relevant
question because an affirmative answer opens the possibility to
predict the  present day values ($\rho_{r0}$, $T_{r0}$) and,
naturally, also the value of the radiation entropy, $S_{r0}$,
associated to the CMB thermal bath.

In principle, the primeval parameters ($\rho_I, H_I$) can be defined
by different approaches. For instance, we may associate directly the
initial de Sitter energy density $\rho_I$ with the Planck
($M_{P}^{4})$ or GUT ($M_X^{4}$) scales. Another possibility  which
will be adopted here is to consider the event horizon (EH) of a pure
de Sitter space-time.  As discussed long ago by Gibbons \& Hawking
(1977), the temperature of the de Sitter EH is $T_{\rm
GH}=(\hbar/k_B)(H/2\pi)$, where $k_B$ is Boltzmann's constant and
$H$ is the (constant) Hubble rate of such initial epoch.  In natural
units, we have ${T}_{GH}={{H}_I}/{2\pi}$. Following this
prescription, and relating $\rho_I$ and $T_{GH}$ according to
Eq.\,(\ref{eq:rhorKolbTurner}), and with the help of (\ref{rhoI}),
we find:

\begin{equation}\label{GibbHI}
{H}_I=\left(\frac{180\pi}{g_*}\right)^{1/2}\,M_{P}\,,\,\,\,\, \ \ \
\ \,\,\,\,\rho_I = \frac{135}{2g*}M_{P}^{4}\,,
\end{equation}
and from Eq.\,(\ref{Temp}) the characteristic radiation temperature
reads:
\begin{equation}\label{GibbTI}
 {T}_{rm}=\left(\frac{45}{2\pi\,g_*}\right)^{1/2}M_{P}\,.
\end{equation}
Therefore, since $g_{*}={\cal O}(100)$ it follows that all primeval
characteristic  scales in the above equations are slightly below the
corresponding Planck scales, and in this sense still within the
classical regime. For example, for $g_{*}=106.75$ we have
${T}_{rm}\simeq 0.26\, M_P$. If we would take into account the
number of light d.o.f in a GUT we would find it even lower,
typically the temperature would be around $10\%$ of the Planck
scale. This means it is still safe to consider a semi-classical QFT
treatment in curved spacetime.

\mysection{Generation of the Radiation Entropy} In natural units,
the radiation entropy contained in the present horizon distance
($d_H\sim H_0^{-1}$) is given by:
\begin{equation}\label{eq:TotalEntropyToday}
S_{0}\simeq \frac{2\pi^2}{45}\,g_{s,0}\,T_{r0}^3\,H_0^{-3}\simeq 2.3
h^{-3} 10^{87}\sim 10^{88}\,.
\end{equation}
Here $T_{r0}\simeq 2.725$$^{\circ}$K $\simeq 2.35\times 10^{-13}$
GeV is the current CMB temperature, $H_0=2.133\,h\times 10^{-42}$
GeV (with $h\simeq 0.67$) is the current Hubble function, and the
coefficient $g_{s,0}=2+6\times
(7/8)\left(T_{\nu,0}/T_{r0}\right)^3\simeq 3.91$ is the entropy
factor for the light d.o.f. today, which involves the well-known
ratio $T_{\nu,0}/T_{r0}=\left(4/11\right)^{1/3}$ of the current
neutrino and photon temperatures (Kolb \& Turner 1990).

\begin{figure}[t]
  \begin{center}
      \resizebox{0.7\textwidth}{!}{\includegraphics{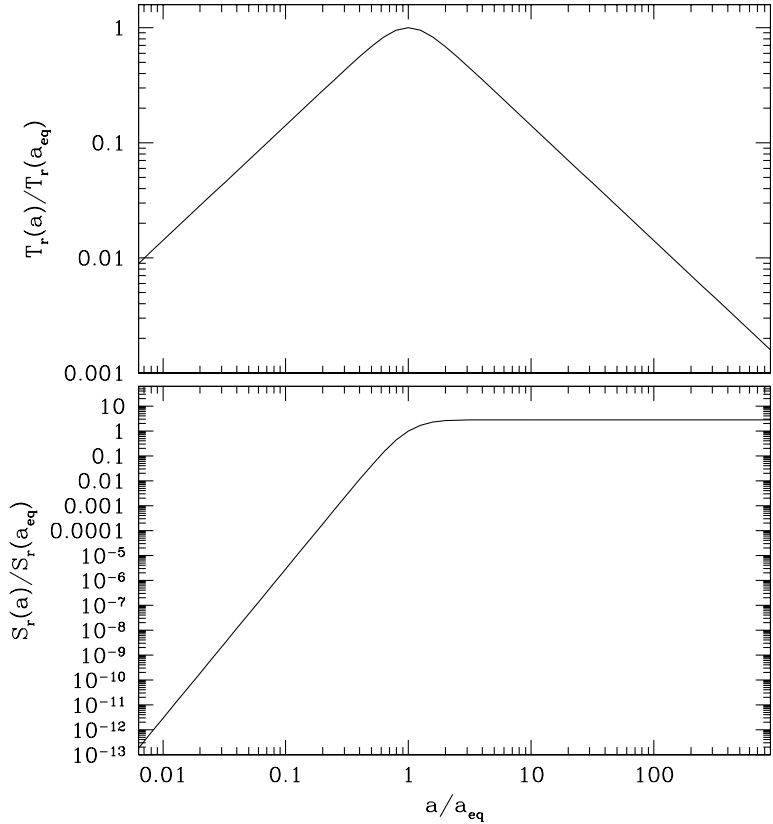}}
      \end{center}
    \caption{{\em Top}:The
evolution of the radiation temperature (normalized with respect to
its maximum value) during the inflationary period and its transition
into the FLRW radiation era. The horizontal axis is as in the
previous figure; {\em Bottom}: The corresponding evolution of the
normalized comoving entropy from the inflationary period (where it
increases) until reaching the saturation plateau for $a/a_{rad} \geq
1$. The asymptotic value corresponds the total entropy at present.}
  \label{Fig2}
\end{figure}

Naturally, if our  model is to be a realistic description of the
cosmic history, the total entropy  measured today should be
computable from the primeval entropy generated by the decaying
vacuum. Let us now check this feature.

Under ``adiabatic" conditions (see Sect. 6), the entropy equilibrium
formula also remains valid because only the temperature law is
modified (Lima 1996, 1997). Therefore, the radiation entropy per
comoving volume reads:
\begin{equation}
S_{r} \equiv \left(\frac{\rho_{r} + p_{r}}{T_{r}}  \right) a^3 =
\frac{4}{3} \frac{\rho_{r}}{T_r} a^3= \frac{2\pi^2}{45} g_{s}
T_{r}^{3} a^3\,,\label{eq19}
\end{equation}
where in the last step we have inserted the expression for the
radiation density, Eq.\,(\ref{eq:rhorKolbTurner}). In the above
equation, $g_s$ is the entropy factor at temperature
$T_r$\,\footnote{Recall that $g_s$ is essentially equal, although
not always identical, to the effective number of massless species,
$g_{*}$. At high temperature this is true for all practical
purposes, but for lower values there is a correction related to the
freeze out of neutrinos and electron-positron annihilation}.

 With the help of equations (\ref{eq19}) and (\ref{eq:Tr}) the comoving entropy can be expressed as a function of the scale factor:
\begin{equation}
S_{r} = \frac{2^{5/2}\pi^2g_{s}}{45}\,T_{rm}^{3}a_{eq}^{3} f(r)\,,
\label{eq:Sr}
\end{equation}
where we have defined the following function of the ratio $r\equiv
a/a_{eq}$:
\begin{equation}\label{eq:deffr}
f(r)=\frac{r^6}{\left(1+r^4\right)^{3/2}}\,.
\end{equation}

Notice that $\lim_{a \rightarrow 0} S_{r} = 0$, as should be
expected since the initial de Sitter state is supported by a pure
vacuum (no radiation fluid). However, during the early stages of the
evolution, deep in the inflationary phase ($a\ll a_{eq}$),  the
total comoving entropy of the Universe increases very fast; in fact,
proportional to $\sim a^{6}$. Note also that for $a=a_{eq}$,
$f(r)=1/\sqrt8$ and the associated value $S(a_{eq})$ is still not
the total comoving  entropy that the decaying vacuum is able to
generate (see discussion below Eq. (\ref{eq3})). This occurs only
when $a=a_{rad}$ so that $r^{4}=(a_{rad}/a_{eq})^{4} \gg 1$ and
$f(r) \simeq 1$ for all practical purposes. At this point  the
comoving entropy $S_r$ saturates to its final value,
$S^{f}_{rad}=S_r(a_{rad})$, thereby marking  the very beginning of
the standard radiation phase.  The evolution of the radiation
entropy normalized to that value is shown in Fig.\,2 (bottom plot),
where the rapid raising of the entropy during inflation and the
final saturation plateau are clearly seen.

It thus follows that the asymptotic adiabatic behavior of
(\ref{eq:Sr}), defined by $f(r) \simeq 1$ for $r\gg1$, is given by:
\begin{equation}\label{eq:Srad1}
S_{r}\to S_{rad}^f=
\left(\frac{2^{5/2}\pi^2g_{s}}{45}\right)T_{rm}^{3}a_{eq}^{3}\,,
\end{equation}
and, as remarked before, such a quantity is the total comoving
entropy predicted by our model. Naturally, its value must be
compared with the present day entropy.

In Fig.\,2 we display the evolution of the entropy as a function of
the ratio $r=a/a_{eq}$, normalized to its value at the
vacuum-radiation equality. Notice that $S_r(a)/S_r(a_{eq})\to
\sqrt{8}$ for $a\gg a_{eq}$.  The plateau characterizes the standard
adiabatic phase, which is sustained until the present days because
the bulk of the vacuum energy already decayed. There is, however, a
remnant vacuum energy parametrized by $c_0$ and $\sim\nu\,H^2$ in
Eq.\,(\ref{powerH}), which is small (recall that
$c_0\sim\rLo\lll\rho_I$ and $|\nu|\lesssim 10^{-3}$). This residual
vacuum energy should describe the current dark energy (DE) epoch
(see LSB for details).

Let us finally check  the prediction of the total entropy within the
current horizon volume $\sim H_0^{-3}$. From the temperature
evolution law (see Eq. (\ref{eq:Tr})) one may check that the product
$T_{rm}^{3}a_{eq}^{3}$ appearing in the final entropy as given by
(\ref{eq:Srad1}) is equal to $2^{-3/2}T_{rad}^{3}a_{rad}^{3}$, where
$T_{rad}=T_r(a_{rad})$ and $a_{rad}$ was defined in Sect. 5. The
result does not depend on the $x$-parameter provided $x\gg 1$, which
holds indeed deep in the radiation epoch. We find

\begin{equation}\label{eq:Srad}
S_{rad}=
\left(\frac{2\pi^2g_{s}}{45}\right){T_{rad}^{3}\,a_{rad}^{3}}=S_0\,,
\end{equation}
where $S_0$ is given by (\ref{eq:TotalEntropyToday}). In the last
step we used the entropy conservation law of the standard adiabatic
radiation phase, which implies that $g_s\,T_{rad}^{3}\,a_{rad}^{3}=
g_{s,0}\,T_{r0}^3\,a_0^{3}$.

In the beginning of the evolution (de Sitter phase), the radiation
entropy is zero.  It increases steadily during inflation but it
asymptotes deep in the radiation epoch.
However, as indicated, the asymptotic value does not depend on the
{\rm x}-parameter used to measure the degree of approximation to the
subsequent phase fully dominated by ultra relativistic particles
i.e., when $q=1$ (see discussion below (\ref{eq15}) and the
associated footnote). Later on, the entropy content evolved to a
plateau  and remained constant till the  observed present day value
(see Fig. 2). The model provides a comprehensible past evolution to
the present $\Lambda$CDM cosmology by evolving smoothly between two
extreme phases driven by the vacuum medium (see Eq.(\ref{HE})). In
particular, the ratio between the associated vacuum energy densities
can be computed using the Gibbons-Hawking ansatz which led to Eq.\,
(\ref{GibbHI}). The final result is:

\begin{equation}
\frac{\rho_{vI}}{\rho_{v0}}=
\frac{H_I^{2}}{H_0^{2}\Omega_{\Lambda_0}}=
\frac{180\pi}{g_*}\,\frac{M_{P}^{2}}{H_0^{2}\Omega_{\Lambda_0}}
\simeq 10^{123},
\end{equation}
 in accordance with usual estimates based on quantum field theory (Weinberg, 1989; Padmanabhan 2003). The model is thus able to explain the radiation entropy content and also to reconcile the observations with the so-called cosmological constant problem. Remarkably, the obtained description of the cosmic history is based on a unified dynamical $\Lambda(H)$-term accounting for both the vacuum energy of the early and of the current universe.

\mysection{Conclusions} In this work we have discussed some
thermodynamical aspects of a non-singular decaying vacuum model (Lima,
Basilakos \& Sol\`a 2013;  Basilakos, Lima \&  Sol\`a 2013)
describing the entire cosmological history of the Universe (namely
from inflation to the current dark energy epoch). The model belongs
to a large class of dynamical vacuum models based on a truncated
series of even powers of the Hubble rate. Such series is
theoretically well motivated within the so-called renormalization
group approach to cosmology (Shapiro \& Sol\`a 2002 and 2009; Sol\`a
2008 and 2013), and hence
carries essential ingredients for a fundamental description of
inflation. Here we have focused on the minimal implementation in which the highest power triggering inflation is $H^4$. The
$H^{2}$ term, which is also present, is neglected for the
description of the early universe, but it can play an important role
for the late time evolution.

The main conclusions of our analysis read as follows:

i) The decaying vacuum cosmological models based on the ``running''
(renormalization group evolution) of the fundamental gravitational
parameters provide a new, qualitatively different, approach to the
description of primordial inflation which is different from the
traditional, inflaton based, mechanisms. The general form of the
decaying vacuum energy density is represented by an even power
series of the Hubble rate that is consistent with the general
covariance of the effective action of quantum field theory in curved
spacetime;

ii) The decaying of the primeval vacuum energy into radiation can
provide a successful account of the ``graceful exit'' from the de
Sitter phase into the standard FLRW radiation epoch, from which the
expansion of the Universe can then follow the concordance $\CC$CDM
scenario;

iii) At the same time we find that the massive decay of the vacuum
energy in this context can explain the large amount of entropy that
we observe in the present Universe. The entropy production can be
explained from a continuous ``heating up'' (rather than the usual
``reheating'' caused by inflaton decay), in which the radiation
temperature rises initially (i.e. during the vacuum phase) linearly
with the scale factor $a$, reaches a maximum at the vacuum-radiation
equality and then decreases inversely proportional to $a$ (as in the
standard adiabatic regime). During the heating up of the Universe
the comoving entropy rockets as the sixth power of $a$ until it
reaches a saturation plateau in the radiation-dominated phase, and
thereafter follows the standard adiabatic regime as well;

iv) Lastly, the decaying vacuum energy is already largely
subdominant during the radiation phase, and therefore the primordial
BBN processes can occur normally. Similarly, the vacuum stays
subdominant during the entire CDM-dominated epoch and the late time
evolution of the Universe can be very close to the $\CC$CDM model,
except for a mild dynamical evolution of the form $\rL(H)\simeq
c_0+c_2 H^2$, which asymptotes to a constant value ($\rL\to c_0$) as
we measure at present.

\vspace{0.2cm}

Let us finally note that, shortly after this work was issued, an
alternative decaying vacuum framework was proposed by (Clifton \&
Barrow 2014) in which the vacuum energy density $\rL$ is also
described in terms of a truncated series of the Hubble rate. It
involves, as in our case, the powers $H^2$ and $H^4$. However, in
their proposal the initial tiny quantum fluctuations are treated as
a manifestation of the thermal nature of quantum fields in a curved
background. Therefore the Universe starts in their case from a state
of radiation with black-body spectrum ($\rho_r\propto T^4$) at a
very high temperature proportional to the Hubble rate (specifically
it is given by the Gibbons-Hawking temperature, $T_{\rm
GH}=H/2\pi$), and as a result $\rho_r\propto H^4$. In our setup we
make also use of the Gibbons-Hawking temperature as a reasonable ansatz, but we assign this temperature to the initial vacuum state rather than
to a primordial radiation heat bath. For this reason the behavior of
the two inflationary pictures is rather different.  In particular,
in contrast to our model, the alternative framework develops a
curvature singularity at early times where both the vacuum energy
and the radiation densities become infinite and with opposite signs
($\rho_r\to +\infty$ and $\rL\to -\infty$ for $t\to 0$). The Clifton
\& Barrow scenario is nevertheless interesting, as it illustrates
with a concrete example that inflation through vacuum decay is
possible even when the energy density of the vacuum is large and
negative. In the particular implementation that they consider, they
find that the very late stage of the universe becomes dominated by
radiation. This is in contradistinction to the situation with our
vacuum decay model, where the asymptotic behavior leads to a
cosmological constant just as in the concordance $\CC$CDM model. It
is remarkable that different inflationary mechanisms based on vacuum
decay dynamics can provide alternative descriptions of inflation which differ substantially from the usual reheating process in conventional
inflationary theories.

\vspace{0.2cm}

In summary, we think it is advisable to keep in mind the mechanisms
of inflation based on vacuum decay. They do not rely directly on the
traditional inflaton fields, but can provide new useful insights on
the inflationary universe that might eventually help to resolve, or
at least highly alleviate, some of the most fundamental cosmological
problems.

\vspace{1cm}


\noindent{\bf Acknowledgments:} \vspace{0.2cm}

JASL is partially supported by CNPq and FAPESP (Brazilian Research
Agencies). JS has been supported in part by FPA2013-46570 (MICINN),
CSD2007-00042 (CPAN) and by 2014-SGR-104 (Generalitat de Catalunya).
SB also acknowledges support by the Research Center for Astronomy of
the Academy of Athens in the context of the program  ``{\it Tracing
the Cosmic Acceleration}''. JASL and SB are also grateful to the
Department ECM (Universitat de Barcelona) for the hospitality and
support when this work was being finished.

\newcommand{\JHEP}[3]{ {JHEP} {#1} (#2)  {#3}}
\newcommand{\NPB}[3]{{ Nucl. Phys. } {\bf B#1} (#2)  {#3}}
\newcommand{\NPPS}[3]{{ Nucl. Phys. Proc. Supp. } {\bf #1} (#2)  {#3}}
\newcommand{\PRD}[3]{{ Phys. Rev. } {\bf D#1} (#2)   {#3}}
\newcommand{\PLB}[3]{{ Phys. Lett. } {\bf B#1} (#2)  {#3}}
\newcommand{\EPJ}[3]{{ Eur. Phys. J } {\bf C#1} (#2)  {#3}}
\newcommand{\PR}[3]{{ Phys. Rep. } {\bf #1} (#2)  {#3}}
\newcommand{\RMP}[3]{{ Rev. Mod. Phys. } {\bf #1} (#2)  {#3}}
\newcommand{\IJMP}[3]{{ Int. J. of Mod. Phys. } {\bf #1} (#2)  {#3}}
\newcommand{\PRL}[3]{{ Phys. Rev. Lett. } {\bf #1} (#2) {#3}}
\newcommand{\ZFP}[3]{{ Zeitsch. f. Physik } {\bf C#1} (#2)  {#3}}
\newcommand{\MPLA}[3]{{ Mod. Phys. Lett. } {\bf A#1} (#2) {#3}}
\newcommand{\CQG}[3]{{ Class. Quant. Grav. } {\bf #1} (#2) {#3}}
\newcommand{\JCAP}[3]{{ JCAP} {\bf#1} (#2)  {#3}}
\newcommand{\APJ}[3]{{ Astrophys. J. } {\bf #1} (#2)  {#3}}
\newcommand{\AMJ}[3]{{ Astronom. J. } {\bf #1} (#2)  {#3}}
\newcommand{\APP}[3]{{ Astropart. Phys. } {\bf #1} (#2)  {#3}}
\newcommand{\AAP}[3]{{ Astron. Astrophys. } {\bf #1} (#2)  {#3}}
\newcommand{\MNRAS}[3]{{ Mon. Not. Roy. Astron. Soc.} {\bf #1} (#2)  {#3}}
\newcommand{\JPA}[3]{{ J. Phys. A: Math. Theor.} {\bf #1} (#2)  {#3}}
\newcommand{\ProgS}[3]{{ Prog. Theor. Phys. Supp.} {\bf #1} (#2)  {#3}}
\newcommand{\APJS}[3]{{ Astrophys. J. Supl.} {\bf #1} (#2)  {#3}}

\newcommand{\Prog}[3]{{ Prog. Theor. Phys.} {\bf #1}  (#2) {#3}}
\newcommand{\IJMPA}[3]{{ Int. J. of Mod. Phys. A} {\bf #1}  {(#2)} {#3}}
\newcommand{\IJMPD}[3]{{ Int. J. of Mod. Phys. D} {\bf #1}  {(#2)} {#3}}
\newcommand{\GRG}[3]{{ Gen. Rel. Grav.} {\bf #1}  {(#2)} {#3}}




{\small
\begin{thebibliography}{}

\bibitem{} Basilakos S., 2009, MNRAS 395, 234; A\&A 508, 575

\bibitem{} Basilakos S., Lima, J. A. S., Sol\`a J., 2013,  Int. J. Mod. Phys. D 22, 1342008 

\bibitem{} Basilakos S., Lima, J. A. S., Sol\`a J., 2014, Int. J. Mod. Phys. D 23, 1442011  


\bibitem{} Basilakos S., Plionis M., Sol\`a J., 2009, Phys. Rev. D.
    80, 3511 

\bibitem{} Basilakos S., Polarski., Sol\`a J., 2012, Phys. Rev. D. 80, 3511 

\bibitem{} Basilakos B.,  Sol\`a J., 2014,  Phys. Rev. D 90, 023008

\bibitem{} Berera A., 1995, Phys. Rev. Lett. 75, 3218


\bibitem{} Carneiro S., Lima J. A. S., 2005, Int. J. Mod. Phys. A,
    20, 2465

\bibitem{} Carvalho J. C., Lima J. A. S.,  Waga I., 1992, Phys. Rev.
    D., 46, 2404

\bibitem{} Carvalho J. C.,  Alcaniz J. S., Lima J. A. S., Silva R.,
    2006, Phys. Rev. Lett. 97, 081301

\bibitem{}  Clifton T., Barrow J. D., 2014,
\textit{Decay of the Cosmic Vacuum Energy}, e-Print:
arXiv:1412.5465.

\bibitem{}  Copeland E.J., Sami M., Tsujikawa S, 2006,  Int. J. Mod.
    Phys. D 15, 1753.

\bibitem{}
 Espa\~na-Bonet C. et al., JCAP, 2004, 0402, 006;
Phys.Lett., 2003, B574, 149




\bibitem{} Fay S., 2014, Phys. Rev. D 89, 063514

\bibitem{} Frampton, P. H., Khepard, T. W., 2008, JCAP 6, 8

\bibitem{} Frampton et al., 2009, Class. Q. Grav. 26, 145005

\bibitem{} Fritzsch H, Sol\`a J., 2012, Class. Quant. Grav.  29, 215002

\bibitem{} Fritzsch H, Sol\`a J., 2014, Adv. High Energy Phys,  2014 361587

\bibitem{} Fritzsch H, Sol\`a J., 2015, \textit{Fundamental constants and cosmic vacuum: the micro and macro connection}, e-Print:  arXiv:1502.01411 (to appear in Mod. Phys. Lett. A).

\bibitem{} Gibbons G. W., Hawking S. W., 1977,   Phys. Rev. D 15,
    2738

\bibitem{}
  G\'omez-Valent A,  Sol\`{a} J., 2014,
  \textit{Vacuum models with a linear and a quadratic term in H: structure formation and number counts analysis}, e-Print: arXiv:1412.3785 (MNRAS, in press).

\bibitem{}
  G\'omez-Valent A., Sol\`{a} J.  Basilakos S., 2015,  JCAP {\bf 01} (2015) 004.


\bibitem{} Grande J., Sol\`a J., Basilakos S., Plonis M., 2011,
JCAP 08, 007

\bibitem{} Guth A. H., Kaiser D. I., Nomura Y,
2014, Phys.Lett., B733, 112 

\bibitem{} Kolb E. W, Turner M. S, \textit{The Early Universe}, 1990, Addison-Wesley Publishing Company.

\bibitem{} Komatsu N., Kimura S., 2014,  Phys. Rev. D {89}, 123501



\bibitem{} Lima J. A. S., 1996, Phys. Rev. D {54},  2571

\bibitem{} Lima J. A. S., 1997, Gen. Rel. Grav. {29}, 805

\bibitem{} Lima J. A. S., 2004, Braz.~Journ.~Phys. {34}, 194, astro-ph/0402109

\bibitem{} Lima J. A. S.,  Baranov I., 2014, Phys. Rev. D 90, 043515

\bibitem{} Lima J. A. S., Basilakos S., Sol\`a J., 2013, MNRAS, 431

\bibitem{} Lima J. A. S.,  Maia J. M. F., 1994, Phys. Rev. D 49, 5597

\bibitem{} Lima J. A. S.,  Trodden M., 1996, Phys. Rev. D 53, 4280

\bibitem{} Linde A, \textit{
Inflationary Cosmology after Planck 2013}, 2013, 100e Ecole d'Ete de
Physique: Post-Planck Cosmology, e-Print: arXiv:1402.0526

\bibitem{} Maia J. M. F.,  Lima J. A. S., 1999, Phys. Rev. D 60, 101301

\bibitem {} Mimoso J. P.,  Pav\'on D., 2013, Phys. Rev. D 87, 047302

\bibitem{} Overduin J. M.,  Cooperstock, F. I., 1998, Phys. Rev. D 58, 043506

\bibitem{} Opher R., Pelinson A., 2004, Phys. Rev. D. 70, 063529

\bibitem{} Padmanabhan T., 2003, Phys.\ Rept. 380, 235

\bibitem{} Peebles P. J. E., 1984, Astrophys. J. 284, 439


\bibitem{} Peebles P. J. E., Ratra, B., 2003, Rev. Mod. Phys.
{75}, 559.

\bibitem{} Perico E. L. D.,  Lima J. A. S., Basilakos  S.,  Sol\`a J., 2013,
Phys. Rev. D. 88, 063531

\bibitem{} Sahni V,  Starobinsky A, 2000, Int. J. of Mod. Phys A9, 373

\bibitem{} Sahni V., Shafieloo A., Starobinsky A. A. 2014

\bibitem{}
Shaefer D. L., Huterer D., 2014,  Phys. Rev. D 89, 063510

\bibitem{} Shapiro I. L.,  Sol\`a, J., 2002,  JHEP 0202, 006

\bibitem{} Shapiro I. L.,  Sol\`a, J., 2009, Phys. Lett. B 682, 105

\bibitem{} Sol\`a J., 2008, J. of Phys. A. Math. Theor. 41, 164066

\bibitem{} Sol\`a J.,  2011, J. Phys. Conf. Ser. 283,  012033 [arXiv:1102.1815]

\bibitem{} Sol\`a J.,  2013, J. Phys. Conf. Ser. 453, 012015 [arXiv:1306.1527].

\bibitem{}  Sol\`a  J., 2014,
Int. J. of Mod. Phys. {A29}, 1444016

\bibitem{} Sol\`a J., 2014, AIP Conf. Proc. 1606, 19  [arXiv:1402.7049]

\bibitem{} Sol\`a J., G\'omez-Valent A., 2015, Int. J. of Mod. Phys. D24 1541003 [e-Print: arXiv:1501.03832]

\bibitem {} Starobinsky,  A. A., 1980,  Phys. Lett. B 91, 99


\bibitem{} 
Steigman, G. 2007, Ann. Rev. Nucl.Part. Sci. 57, 463 [e-Print:
arXiv:0712.1100]

\bibitem{}  Vilenkin A., 1982, Phys. Lett.  B 117, 25

\bibitem{} Weinberg S., 1989, Rev.\ Mod.\ Phys. 61, 1



\end{thebibliography}}
\end{document}